\documentclass[manuscript]{aastex}

\usepackage{graphics,color}
\usepackage{natbib}
\definecolor{dark}{gray}{0.5}
\definecolor{red}{rgb}{1,0,0}
\definecolor{green}{rgb}{0,1,0}
\definecolor{blue}{rgb}{0,0,1}

\shorttitle{The Bimodal Color Distributions Of The Milky Way From 2MASS SC}
\shortauthors{Chang et al.}
\begin{document}
\title{The Information Of The Milky Way From
2MASS Whole Sky Star Count: The Bimodal Color Distributions}

\author{Chan-Kao Chang \altaffilmark{1}}
\affil{Institute of Astronomy, National Central University, Jhongli, Taiwan}

\author{Shao-Yu Lai}
\affil{Institute of Astronomy, National Central University, Jhongli, Taiwan}

\author{Chung-Ming Ko \altaffilmark{2}}
\affil{Institute of Astronomy, Department of Physics and Center of Complex Systems, \\
National Central University, Jhongli, Taiwan}

 \and
\author{Ting-Hung Peng}
\affil{Institute of Astronomy, National Central University, Jhongli, Taiwan}

\altaffiltext{1}{rex@astro.ncu.edu.tw} \altaffiltext{2}{cmko@astro.ncu.edu.tw}

\begin{abstract}
The $J-K_s$ color distribution (CD) with a bin size of 0.05 magnitude for the
entire Milky Way has been carried out by using the Two Micron All Sky Survey
Point Source Catalog (2MASS PSC). The CDs are bimodal, which has a red peak at
$0.8 < J-K_s < 0.85$ and a blue peak at $0.3 < J-K_s < 0.4$. The colors of the
red peak are more or less the same for the whole sky, but that of the blue peak
depend on Galactic latitude ($J-K_s \sim 0.35$ at low Galactic latitudes and
$0.35 < J-K_s < 0.4$ for other sky areas). The blue peak dominates the bimodal
CDs at low Galactic latitudes and becomes comparable with the red peak in other
sky regions. In order to explain the bimodal distribution and the global trend
shown by the all sky 2MASS CDs, we assemble an empirical HR diagram, which is
composed by observational-based near infrared HR diagrams and color magnitude
diagrams, and incorporate a Milky Way model. In the empirical HR diagram, the
main sequence stars turnoff of the thin disk is relatively bluer,
$(J-K_s)_0=0.31$, when we compare with the thick disk which is
$(J-K_s)_0=0.39$. The age of the thin/thick disk is roughly estimated to be
around 4-5/8-9 Gyr according to the color-age relation of the main sequence
turnoff. In general, the 2MASS CDs can be treated as a tool to census the age
of stellar population of the Milky Way in a statistical manner and to our
knowledge this is a first attempt to measure the age.
\end{abstract}


\section{Introduction}
One of the obvious properties of the Milky Way that we want to learn is its
morphology. The fact that we reside inside the Milky Way hinders us from a
global view of our own Galaxy which we enjoy over other galaxies. William
Herschel made use of star count (SC) to get a rough idea of how the Milky Way
looks like in the eighteenth century \citep{Hers1785}. The advance in data
collecting and analysis methods in late twentieth century let research groups
\citep[such as][]{Bahcall1980, Gilmore1985} to quantify the major components
(i.e., disks and halo) of our Galaxy by SC. They concluded that the properties
of the stellar population in different components are distinguishable
\citep[see Table 1 in ][and references therein]{Chang2011}. Although many
studies have been carried out, a common consensus on the structural parameters
of the Milky Way has not been reached yet. One of the reasons can be attributed
to different sky regions and limiting magnitudes (i.e., limiting volumes) were
used \citep{Siegel2002, Karaali2004, Bilir2006a, Bilir2006b, Karaali2007,
Juric2008}. This problem can be overcome by wide sky coverage SC studies which
become possible in the last decade as modern all sky surveys become available,
such as the Two Micron All Sky Survey \citep[2MASS;][]{Skrutskie2006}, the
Sloan Digital Sky Survey \citep[SDSS;][]{York2000}, the Panoramic Survey
Telescope \& Rapid Response System \citep[Pan-Starrs;][]{Kaiser2002} and the
Wide-field Infrared Survey Explorer \citep[WISE;][]{Wright2010}. However,
subsequent researches did not settle the issue and the variation in different
studies is thought to come from the degeneracy between the structural
parameters. Moreover, the Galactic structural parameters have been reported to
show dependence on Galactic longitude and latitude \citep{Bilir2006a, Ak2007,
Cabrera2007, Bilir2008, Yaz2010}. This makes the measurement and interpretation
far more complex and challenging.

To solve the degeneracy, information other than SC is needed, e.g., dynamics,
color, etc. The most handy tool is stellar color as it is available in all sky
surveys. That is the reason why most SC studies come with color distribution
(CD), which calculates the number of star in each color bin. In addition to
luminosity function (LF) and density profile (DP) used in single wavelength SC
study, CD requires Hertzsprung-Russell diagram (HR diagram) to transform each
luminous star into its corresponding color. HR diagram could be empirical
(i.e., from observation) or theoretical (i.e., from population syntheses). In
the empirical case, HR diagram and number densities of each spectral type star
are provided from stellar cluster and local field stars \citep[see
e.g.,][]{Bahcall1980, Mamon1982, Wainscoat1992}. Difficulties include
time-consuming data collection and processing, and possible bias of sample
selection (e.g., incompleteness, improper representation). In general,
observational HR diagram in optical wavelength is obtained first and then other
waveband HR diagrams are constructed by color transformation \citep[see][for
examples]{Mamon1982,Wainscoat1992}. Thus, the uncertainty in color
transformation needs to be considered carefully. In the theoretical case, HR
diagram is deduced from the Hess diagram, which is generated by population
synthesis based on knowledge of the initial mass function, stellar evolution
and star formation history, etc. The Besan\c{c}on model \citep{Robin2003} and
the TRILEGAL model \citep{Girardi2005} are examples. Moreover, population
synthesis models depend greatly on input parameters, and stellar evolution and
atmosphere.

In \citet{Chang2010, Chang2011}, we have shown that the $K_s$ all sky 2MASS SCs
can be well described by a single power law LF and a three-component DP
(namely, the thin disk, the thick disk and the halo). In this work, we would
like to extend our model to explain the $J-K_s$ CDs of the entire Milky Way. We
gather currently available NIR data, i.e., from direct NIR observations,
transformations of optical HR diagram, and observational NIR color magnitude
diagram (CMD) of star clusters, to assemble an empirical NIR HR diagram and
incorporate a Milky Way model to explain the 2MASS data. In a way, the HR
diagram we used is a 2MASS optimized HR diagram. The article is organized as
follows. The features of the all sky 2MASS CD are described in section 2. Our
Milky Way model, including LF and DP, is provided in section 3. The procedure
to create our Hess diagram and HR diagram is given in section 4. In section 5
we present our results on CDs of the Milky Way, and section 6 is a summary and
concluding remarks.

\section{The 2MASS Data and Its Color Distributions}
We use the 2MASS Point Source Catalog \citep[2MASS PSC,][]{Cutri2003} to carry
out $J-K_s$ CDs with a bin size of 0.05 magnitude for the entire Milky Way. We
select objects with the following criteria: (1) signal-to-noise ratio $\ge 5$,
(2) detection in all $J$, $H$, $K_s$ bands and (3) $K_s$ magnitude between 5
and 14 mag. The last criterion ensures 99\% completeness rate before the $K_s$
limiting magnitude (i.e., 14.3 mag) and avoids the relatively large photometric
error for $K_s \le 5$ mag objects. The whole sky is divided into 8192 nodes
according to level 5 Hierarchical Triangular Mesh \citep[HTM,][]{Kunszt2001},
which samples the whole sky roughly evenly and has a 2 degrees angular
separation on average. The radius of each node is 1 degree (i.e., each node
covers $\pi$ square degree). Because of the shallower limiting magnitudes and
the complex extinctions in the Galactic center area, this work does not attempt
to explain the 2MASS CDs in this region.

\begin{figure}
  \epsscale{0.8}
  \plotone{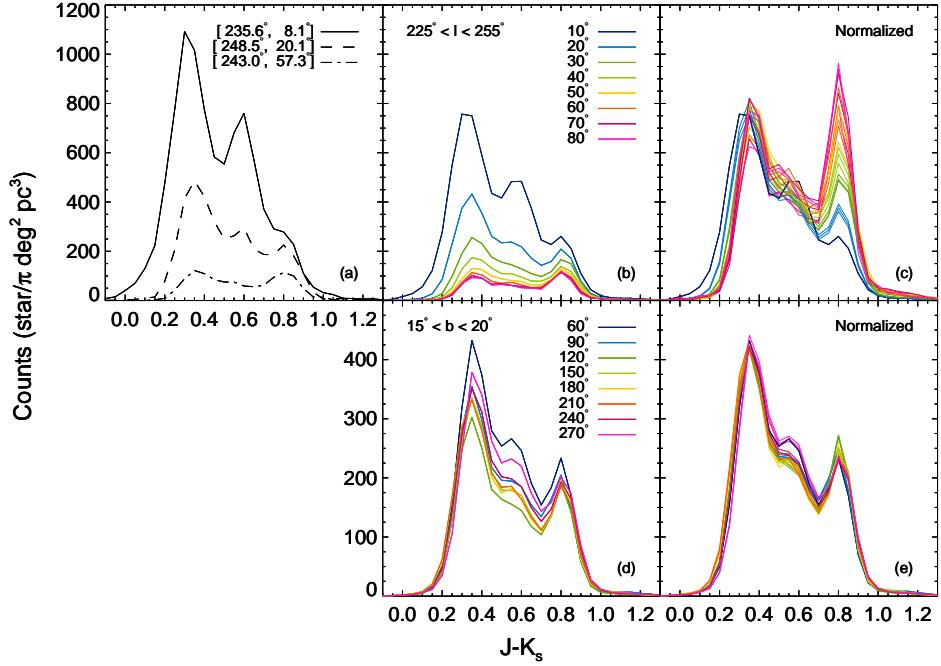}
  \caption{Salient features on the 2MASS CD. (a) Typical examples of 2MASS CD,
  $[l,b]=[235.6^\circ, 8.1^\circ]$ (solid line), $[248.5^\circ, 20.1^\circ]$ (dashed line) and
  $[243.0^\circ, 57.3^\circ]$ (dotted-dashed line).
  (b) The 2MASS CDs averaged over $225^{\circ} < l < 255^{\circ}$ at different Galactic latitudes.
      The Galactic latitudes are indicated by different colors.
  (c) Same as (b), except that the total count is normalized to that of $b = 10^{\circ}$.
  (d) The 2MASS CDs averaged over $15^{\circ} < b < 20^{\circ}$ at different Galactic longitudes.
      The Galactic latitudes are indicated by different colors.
  (e) Same as (d), except that the total count is normalized to that of $l = 60^{\circ}$.}
  \label{2MASSCD}
  \end{figure}

\begin{figure}
  \epsscale{0.8}
  \plotone{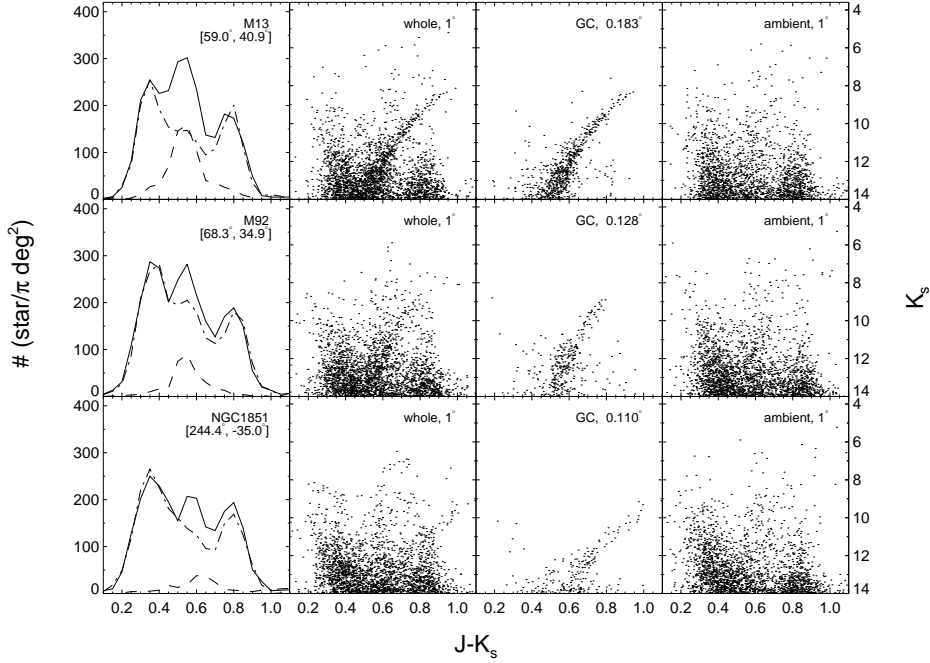}
  \caption{The first column is the 2MASS CDs having the extra peak at medium and high Galactic latitudes.
  The whole field, the globular-cluster-only and the ambient field CDs are represented by the solid, dashed
  and dotted-dashed lines, respectively. The identification of the star cluster and its center coordinate
  are indicated on the upper-right corner of each figure.
  The second, third and last columns show the CMDs of the whole field, the globular-cluster-only and the
  ambient field, respectively. The radii of the field of view of the whole field and the ambient field CMDs
  are 1 degree and that of globular-cluster-only CMD is given on the upper-right corner of each figure in the
  third column.}
  \label{alter_peak}
  \end{figure}

\begin{figure}
  \epsscale{0.8}
  \plotone{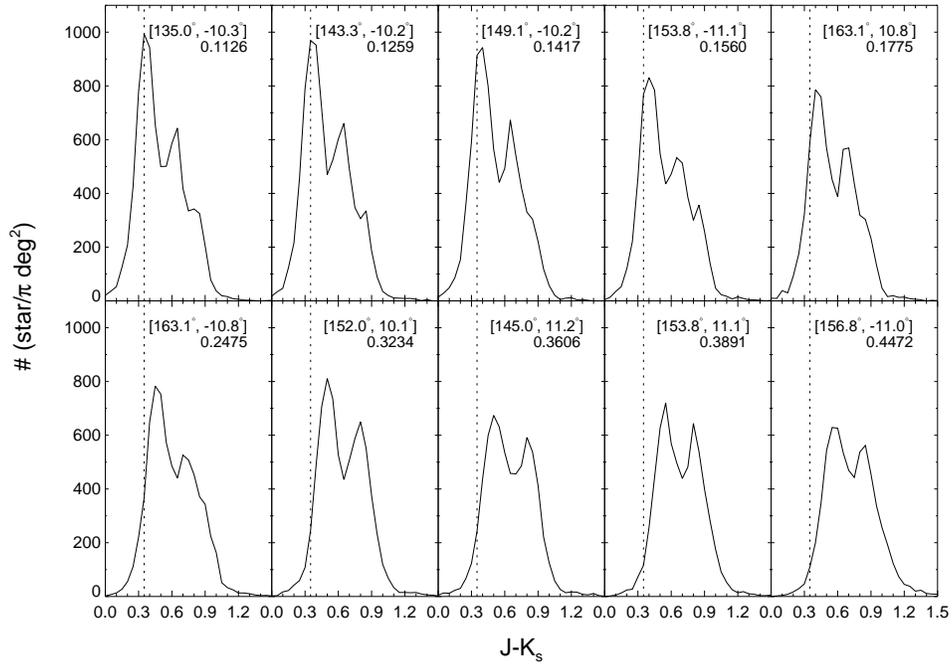}
  \caption{The 2MASS CDs around $|b|\sim 16^\circ$ with different $E(J-K_s)$ extinction values.
  The legend gives the Galactic coordinates and the $E(J-K_s)$ value.}
  \label{extshape}
  \end{figure}

Several common features are identified in the whole sky 2MASS CDs, which are
listed below and demonstrated in Fig.~\ref{2MASSCD}.
\begin{enumerate}\label{CDFeatures}
\item
Most of the 2MASS CDs are bimodal, which has a blue peak at $0.3 < J-K_s <
0.4$ and a red peak at $0.8 < J-K_s < 0.85$. The 2MASS CDs at low Galactic
latitudes have one more peak at $0.55 < J-K_s < 0.6$. Fig.~\ref{2MASSCD}a
shows three examples of typical 2MASS CD at different Galactic latitudes.

\item
The blue peak dominates the 2MASS CD at low Galactic latitudes and becomes
comparable to the red peak at high Galactic latitudes. The middle peak only
shows up at low Galactic latitudes and does not exist in other sky areas
any more. Fig.~\ref{2MASSCD}b shows the averaged CDs over $225^{\circ} < l
< 255^{\circ}$ at different Galactic latitudes.

\item
The $J-K_s$ colors of the all sky red peaks and the middle peaks at low
Galactic latitudes are almost fixed. However, the colors of the blue peak
depend slightly on Galactic latitudes, which is $J-K_s \sim 0.35$ at low
Galactic latitudes and $0.35 < J-K_s < 0.4$ at medium and high Galactic
latitudes. The situation is demonstrated in Fig.~\ref{2MASSCD}c, which is
the same as Fig.~\ref{2MASSCD}b except that the total count is normalized
to $b = 10^{\circ}$. In contrast, the color of the blue peak does not
change along Galactic longitude, see Fig.~\ref{2MASSCD}d \& \ref{2MASSCD}e.

\item
We call the tapering off distribution on the blue side of the blue peak the
blue wing. The shapes of the blue wings at different Galactic latitudes are
very similar. Due to their bluer blue peaks, the blue wings of the CD at
low Galactic latitudes can extend to $J-K_s = 0$. However, the blue wings
at medium and high Galactic latitudes seldom extend to less than $J-K_s =
0.2$. The normalized 2MASS CDs in Fig.~\ref{2MASSCD}c clearly shows this
property.

\item
Occasionally, some 2MASS CDs at medium or high Galactic latitudes have a
middle peak, which is different from the middle peak at low Galactic
latitudes. We call it `the extra peak' and will explaine more in
section~\ref{sec1} and \ref{sec2}. The solid lines in the first column of
Fig.~\ref{alter_peak} show three examples of the extra peak.

\item
The shape of the 2MASS CD can be altered significantly by severe
extinction, which is the usual case at low Galactic latitudes.
Fig.~\ref{extshape} shows the 2MASS CDs at $|b| \sim 16^{\circ}$. The
figures are arranged according to their $E(J-K_s)$ extinction values. As
the $E(J-K_s)$ values increase, the sharpness of the 2MASS CDs decrease
gradually, and the $J-K_s$ colors of the blue and middle peaks become
redder and redder. Moreover, the tail in the red end is elongated.
\end{enumerate}

\section{The Milky Way Model}
Our Milky Way model has a three-component DP and a single power law LF
\citep{Chang2010, Chang2011}. The three-component DP $n(R,Z)$ includes a thin
disk $D_1$, a thick disk $D_2$ and an oblate halo $S$,
\begin{equation}\label{density}
  n(R,Z)=n_0\left[D_1(R,Z)+D_2(R,Z)+S(R,Z)\right]\,,
\end{equation}
where $R$ is the galactocentric distance on the Galactic plane, $Z$ is the
distance from the Galactic mid-plane and $n_0$ is the local stellar density of
the thin disk at the solar neighborhood.

The disks are in a double exponential decay form in which stellar density
decreases exponentially along $R$ and $Z$,
\begin{equation}\label{Di}
  D_i(R,Z)=f_i\exp\left[-\,{(R-R_\odot)\over H_{ri}}-\,{(|Z|-|Z_\odot|)\over H_{zi}}\right]\,,
\end{equation}
where $(R_\odot,Z_\odot)$ is the location of the Sun, $H_{ri}$ is the
scale-length, $H_{zi}$ is the scale-height, and $f_i$ is the density ratio to
the thin disk at the solar neighborhood. The subscript $i=1$ stands for the
thin disk and $i=2$ stands for the thick disk.

The halo is a power law decay oblate spheroid flattening in the $Z$ direction,
\begin{equation}\label{SH}
  S(R,Z)=f_h\left[R^2+(Z/\kappa)^2\over R_\odot^2+(Z_\odot/\kappa)^2\right]^{-p/2}\,,
\end{equation}
where $\kappa$ is the axis ratio, $p$ is the power index and $f_h$ is the local
halo-to-thin disk density ratio.

The $M_{K_s}$ LF is a single power law \citep{Chang2010},
\begin{equation}\label{LF}
  \psi(M_{K_s})={2\log_{\rm e}10\ (\gamma-1)\over
  5\left[10^{2(\gamma-1)M_{f}/5}-10^{2(\gamma-1)M_{b}/5}\right]}\,10^{2(\gamma-1)M/5}\,,
\end{equation}
where $\gamma$ is the power law index, $M_b$ and $M_f$ are the bright and faint
cutoffs, respectively. Note that $\psi(M)$ includes all luminosity classes. For
convenience, we adopt the peak values of the bright end and faint end
distributions of \citet{Chang2010}, i.e., $M_b=-8$ and $M_f=6.5$ \citep[see
Fig.~6 in][]{Chang2010}, as the bright and faint ends in this study. We note
that the averages in \citet{Chang2010} are $M_b=-7.86\pm 0.60$ and $M_f=6.88\pm
0.66$. Nevertheless, the model prediction shows no significant difference
between using the peak values and the average ones.

The extinction model is adopted from the new COBE/IRAS result \citep{Chen1999}
and the color excess is $E(J-K_s)/E(B-V)=0.53$ \citep{Schlegel1998}.
Table~\ref{model} lists the parameters taken from \citet[][]{Chang2010} and
\citet[][]{Chang2011} (hereafter MW-I).

\section{The Empirical NIR HR Diagram Optimized for the 2MASS CD}\label{diagram}
We describe the search for an empirical NIR HR diagram (tailored for 2MASS
data) in three subsections for a simple and clear exposition.

\subsection{Basic NIR HR Diagram}\label{sec1}
We take observational-based NIR HR diagrams in the literature as our first
construction of the loci of main sequence (MS) and giant branch (GB) in
$M_{K_s}$-$(J-K_s)_0$ HR diagram. \citet[][hereafter W92]{Wainscoat1992} and
\citet[][hereafter C07]{Covey2007} have the information of NIR HR diagram
(i.e., the NIR color, absolute magnitude and spectral type). W92 also gives the
number density in the solar neighborhood and the normalization in different
components of Milky Way. Fig.~\ref{HRD} shows the data points of MS, GB and
super giants taken from W92 and C07. Both data sets agree to each other very
well and shows several common features: (1) a sharp downturn around the end of
MS (hereafter, MS-downturn), (2) a slight upturn around late A type star (i.e.,
$M_{K_s} \sim 3$ mag and hereafter, MS-turnoff), and (3) a sharp curve on GB
around $M_{K_s} \sim -1$ mag (hereafter, GB-curve). We fit W92 data to find the
loci and integrate it with MW-I to obtain a synthetic CD. When the synthetic CD
is compared with 2MASS CD, some significant discrepancy is observed: (1) over
predictions for the middle peak and for the blue wing (i.e., $J-K_s < 0.3$
mag), (2) under prediction on the blue peak, and (3) the locations of the blue
and middle peaks are not correct. The first column in Fig.~\ref{SyntheticCD}
demonstrates several results for different Galactic latitudes. However, we
obtain a very useful information that the $J-K_s$ colors of each peak
correspond to that of each sharp turn (or `vertical' sections) on the loci.
Apparently, the blue, red and middle peaks are the results of the MS-turnoff
around late A type stars, the MS-downturn and the GB-curve, respectively.

\subsection{The Optimization for the Empirical NIR HR Diagram}\label{sec2}
To study the sharp turns in more detail, we utilize several NIR CMDs of star
clusters from the literature, such as \citet[][]{Beletsky2009, Sarajedini2009,
Troisi2010}. Some of them \citep[e.g., M67 and NGC6791 in][]{Beletsky2009,
Sarajedini2009}, which have information down to the faint end of MS, are used to
conclude the important features for our analysis: (1) the MS-downturn
$(J-K_s)_0 \sim 0.83$ is a common feature at the faint end of MS; (2) the
MS-turnoff (i.e., the slight upturn around late A star) reflects stars
departing from the MS at the end of their MS stage; (3) the MS-turnoff always
accompanies a $\Delta M_{K_s} \sim 2.5~K_s$ mag vertical height extension and
its $(J-K_s)_0$ color depends on the age of the star cluster; and (4) the main
sequence belt between the MS-turnoff and the MS-downturn is approximately
linear. Since only few GB stars are shown on these NIR CMDs, we are not able to
get more details of the GB-curve. However, we believe that the location of the
GB-curve should be similar to that of W92 and C07.

Accordingly, a working HR diagram includes the following features. The MS is
divided into three parts: (1) early MS (i.e., O-A type stars and $M_{K_s}
\lesssim 3$ mag); (2) nearly linear late MS (i.e., F-K type stars and $3
\lesssim M_{K_s} \lesssim 5$ mag); and (3) sharp and almost vertical
MS-downturn at $(J-K_s)_0 \sim 0.83$ (i.e., M type and later stars and $M_{K_s}
\gtrsim 5$ mag). The GB is separated into up and low GBs. Both are linear and
connected by the GB-curve at $(J-K_s)_0 \sim 0.6$ and $M_{K_s} \sim -1$.
Consequently, we create an empirical $M_{K_s}$ against $(J-K_s)_0$ locus
including a vertical MS-downturn, a linear late MS, a MS-turnoff with a $\Delta
M_{K_s} = 2.5~K_s$ mag vertical extension, a linear low GB, a linear up GB and
a flat transition between the tip of the MS-turnoff and the low GB. The slopes
of each linear part (i.e., the up GB, the low GB and the late MS) and the
connecting joints of different parts are taken from the CMDs mentioned above
and the data in W92 and C07. The early MS is not included because the 2MASS CDs
are generally deficient in number of stars in the region $(J-K_s)_0 < 0.2$.
Since the $J-K_s$ colors of the red peak of 2MASS CD is between 0.8 to 0.85 and
that of the MS-downturns of M67 and NGC6791 are at $\sim 0.83$
\citep{Beletsky2009, Sarajedini2009}, we set $(J-K_s)_0 = 0.83$ for the
vertical MS-downturn. Moreover, the middle peak is around $J-K_s=0.55$ to 0.6
and the GB-curves in W92 and C07 also have similar color, so we set $(J-K_s)_0
= 0.6$ for the GB-curve and let the low GB between $(J-K_s)_0 = 0.5$ and 0.6.
To incorporate the idea that different $(J-K_s)_0$ colors of the MS-turnoff for
different evolutionary stage populations \citep{Sarajedini2009}, we allow the
MS-turnoff to move along the main sequence belt. The left panel of
Fig.~\ref{HRD} shows the empirical NIR locus on top of the data points taken
from W92, C07 and M67 \citep{Beletsky2009}. To mimic observational color
scattering, we put a 0.06 color dispersion to the empirical locus (see the
right panel of Fig.~\ref{HRD}).

Incorporating this empirical NIR HR diagram with MW-I, we found several
features in the model prediction (see Fig.~\ref{SyntheticCD}): (1) the CD
generated by the thin disk is a bimodal distribution; (2) the CD generated by
the thick disk only shows the blue peak; (3) both CDs of the thin and thick
disk can have a middle peak at low Galactic latitudes; (4) the overall model
prediction is mainly dominated by the CD of the thin disk; (5) the blue peak of
the thick disk has an appreciable contribution to that of the overall model
prediction only at medium and high Galactic latitudes; and (6) the red peak of
the overall model prediction is determined mostly by the thin disk.

At first, we focus on medium and high Galactic latitudes to avoid severe
extinction. The $(J-K_s)_0$ colors of the MS-turnoff for the thin and thick
disk are set between 0.35 and 0.4 to comply with the blue peak of 2MASS CD at
medium and high latitudes. We integrate the empirical locus (Fig.~\ref{HRD})
with our Milky Way model MW-I to generate the synthetic CD. For reasonable good
fit, the $(J-K_s)_0$ color of the thin disk and the thick disk are found to be
0.34-0.35 and 0.38-0.39, respectively. The second column in
Fig.~\ref{SyntheticCD} shows some results at different Galactic latitudes.

The result at medium and high Galactic latitudes agrees well with the 2MASS
data. Then, we inspect low Galactic latitudes which have relatively small
extinction. The result shows significant difference in the blue wing of the
blue peak (i.e., $J-K_s < 0.3$), but the rest of the CD is still in a good
agreement (see the first row of the second column in Fig.~\ref{SyntheticCD}).
If a bluer turnoff is assigned to the thin disk, the discrepancy in the blue
wing can be compensated somewhat, but it will cause an over prediction in the
blue wing and a shift in $J-K_s$ color of the blue peak at high Galactic
latitudes (see the third column in Fig.~\ref{SyntheticCD}).

\subsection{The Final Adjustment}\label{sec3}
When experimenting parameters of the model in previous subsections, we notice a
small scale height thin disk would make its blue peak change more dramatically
along Galactic latitude. This makes the overall blue peak (i.e., the thin and
thick disks ) at low Galactic latitudes be dominated by the thin disk, while
the blue peak at medium and high Galactic latitudes be dominated by the thick
disk. Therefore, a small scale height thin disk with a bluer MS-turnoff not
only can eliminate the difference in the blue wing at low Galactic latitudes,
but also can reduce its contribution to the blue peak at medium and high
Galactic latitudes. In this way, the $J-K_s$ color of the blue peak could be
relatively bluer at low Galactic latitudes and relatively redder at the other
sky region. With this strategy, we select an acceptable Galactic structure
configuration with a relatively smaller scale height thin disk from our
previous SC study \citep[see][for details]{Chang2011} and come up with the best
fit model as follows: (1) the scale height of the thin disk is $H_{z1}=260$ pc
and other structural parameters are listed in Table~\ref{model} (hereafter,
MW-II); and (2) the MS-turnoff of the thin disk is $(J-K_s)_0=0.31$ mag. With
this model, the 2MASS CDs at low Galactic latitudes can be fitted much better
than that in subsection~\ref{sec2} (see the last column in
Fig.~\ref{SyntheticCD}).

Fig.~\ref{HRD} shows the HR diagram obtained from subsections~\ref{sec2} and
\ref{sec3}. Incorporating the luminosity function derived from 2MASS SC
\citep{Chang2010}, we construct the Hess diagram of MW-II in the right panel of
Fig.~\ref{HRD}. The joints on the HR diagram of MW-II are listed in
Table~\ref{THRD}.

\section{Results and Discussion}\label{results}
In Figs.~\ref{result1} \& \ref{result2} we compare our model prediction (i.e.,
MW-II in section~\ref{sec3}) and the 2MASS CDs at different Galactic latitudes.
In order to minimize the uncertainty caused by extinction correction at low
Galactic latitudes, we compare sky areas with relatively small extinction. The
model prediction and the 2MASS CD agree very well in the overall shape and the
$J-K_s$ color of each peak. The number variations along Galactic latitudes of
each peak are well reproduced by the model.

In order to present the global trend of the 2MASS CDs, we show the number ratio
of the blue part (i.e., $J-K_s \leq 0.6$) to the red part (i.e., $J-K_s > 0.6$)
in Fig.~\ref{bridx} along with the prediction of MW-II. Both figures have very
similar trend which is relatively blue at low Galactic latitudes and become
comparable in other sky regions. The very red Galactic disk is the result of
severe extinction.

When we inspect the Galactic disk region (i.e., $|b| < 10^{\circ}$),
the model prediction has obvious discrepancy with the 2MASS CD. The discrepancy
can mostly be attributed to improper extinction correction and number
inconsistency between the model prediction and the 2MASS data. The number
inconsistency, which has been reported in \citet{Chang2011}, results in gaps on
each peak between the model prediction and the 2MASS CD. Fig.~\ref{inconsist}
demonstrates some cases together with the SC results taken from
\citet{Chang2011}. In these cases, the sky areas have relatively small
extinction. This helps us to delineate the discrepancy mainly comes from the
number inconsistency (see the difference in SC result, lower panels in
Fig.~\ref{inconsist}). Several examples of improper extinction correction are
illustrated in the first column of Fig.~\ref{bluep}. The improper extinction
can make a significant offset on the $J-K_s$ colors of the blue and middle
peaks and distort the shape of CD. Moreover, we also find a significant blue
population (i.e., $J-K_s <0.2$) in the 2MASS CD on the Galactic disk ($|b| <
10^{\circ}$) around the Galactic longitudes of $180^{\circ} < l < 240^{\circ}$
(see the second column of Fig.~\ref{bluep}). This blue population is neither
predicted by the model nor the blue wing of the blue peak. If we omitted the
extinction correction, the model still cannot account for such a blue
population. The origin of this blue population warrants further studies.

Several 2MASS CDs at medium and high Galactic latitudes have an extra peak at
$J-K_s \sim 0.55$. Their occurrence and height do not correlate with Galactic
coordinates. Most of them are insignificant, but we show three good examples in
the first column of Fig.~\ref{alter_peak} (solid line). We note that each
example has a globular cluster within the field of view. In order to nail down
the origin of the extra peak, we generate globular-cluster-only CD (dashed
line). Each globular-cluster-only CD has a spike, which has $J-K_s$ color very
similar to that of the extra peak in the whole field CD (i.e., the CD contains
stars both from the globular cluster and the field; solid line). We also obtain
the CD of the ambient field (dot-dashed line) by subtracting the
globular-cluster-only CD from the whole field CD. The extra peak does not show
up anymore in the ambient field CD. Besides, each globular cluster CMD has an
obvious GB with an obscure GB-curve at $J-K_s \sim 0.55$ similar to that on the
empirical HR diagram (see Fig.~\ref{HRD}). Therefore, we conclude that the
extra peak is due to the low GB of the globular cluster and its significance
depends on the relative amount of star in the low GB of the globular cluster to
that in the ambient field. Besides, if the limiting magnitude could be reached
down to the MS-turnoff of the globular clusters, the observed blue peak would
have more stars than the model prediction. Similar number enhancement could
happen in the 2MASS CDs if there is an overdensity within the field of view,
which have a dominant stellar population over the smooth stellar distribution
of the Milky Way (e.g., star cluster, stellar stream, Galactic bulge, Galactic
arm, etc.). For instance, the extremely dense Galactic bulge, which is not
included in our model, makes a prominent middle peak on the 2MASS CDs, which
obviously outnumbers the model prediction (see the third column of
Fig.~\ref{bluep}). However, we do not observe any obvious number enhancement in
other 2MASS CDs which have known stellar streams in their field of view, e.g.,
Sagittarius dwarf galaxy \citep{Majewski2003}, Monoceros stream \citep[][;
Table 1]{Penarrubia2005}, etc. It is possible that within the 2MASS detection
limit the populations of these overdensities are too small.

At low Galactic latitudes there is another possible way to generate the middle
peak. It is the red clump stars \citep{Alves2000, Cabrera2007}, which occupies
the space close to the low GB on our HR diagram and has a higher number density
relative to the other stars on GB. Such relatively higher number density would
produces more stars in the corresponding color bins. In our model we do not
include the red clump stars in our LF and HR diagram, but our model still have
good agreement with the 2MASS CD. The reason might be the net contribution of
the red clump stars in generating the middle peak is similar to that of the low
GB.

Our best model that can explain the all sky 2MASS CDs (see section~\ref{sec3})
comprises a thin disk, which has a relative short scale height (260 pc) and a
relative blue MS-turnoff (i.e., younger), and a thick disk (1040 pc), which has
a relative red MS-turnoff (i.e., older). Our best model, MW-II, that can
explain the all sky 2MASS CDs (see section~\ref{sec3}) comprises a thin disk
(260 pc) with a relatively blue MS-turnoff (0.31 mag, i.e., younger), and a
thick disk (1040 pc) with a relatively red MS-turnoff (0.38 mag, i.e., older).
The two populations can be interpreted in the context of the formation of the
Milky Way. The thick disk was formed in an epoch earlier than the thin disk.
Moreover, both disks have the same single power law LF which implies that the
initial mass function does not change along the Milky Way's evolution (unless
the mass-luminosity relation was time dependent in the past). In this sense,
the all sky 2MASS CDs can be used as a tool to census the age of stellar
population in the Milky Way. If we apply the color-age relation of MS-turnoff
of \citet{Sarajedini2009} to our empirical HR diagrams, we can roughly estimate
the ages of the thin and thick disk to be 4-5 Gyr and 8-9 Gyr, respectively.
The age estimation should be viewed as the lower limit. The blue boundary of
the 2MASS CDs do not allow bluer MS-turnoffs of the disks. Moreover, the 0.05
mag bin size of our 2MASS CD can only provide a low color resolution of the
blue peak, hence the color of the MS-turnoff and the derived age cannot be
estimated more precisely. Nevertheless, to the best of our knowledge our result
could be the first attempt to measure the age of global stellar components in a
general statistical sense. We conclude that the whole population of the thick
disk is about 3-4 Gyr older than that of the thin disk.

In this work, we seldom mention the halo component because it has very limited
contribution to the total CDs in magnitude range we are interested (i.e., $4\
{\rm mag}< K_s < 14\ {\rm mag}$). Thus, we are not able to extract solid and
useful information of the halo from 2MASS CD.

\section{Summary}
We use 2MASS PSC to carry out the whole sky $J-K_s$ CDs, which show a bimodal
distribution with a blue peak and a red peak. The blue peak is the dominant
feature at low Galactic latitudes and becomes comparable to the red peak at
high Galactic latitudes. The all sky $J-K_s$ colors of the red peak are almost
the same (between 0.8 and 0.85 mag). However, the $J-K_s$ colors of the blue
peak at low Galactic latitudes are a bit bluer (0.35 mag) than that at medium
and high Galactic latitudes (between 0.35 mag and 0.4 mag). However, the
$J-K_s$ colors of the blue peak at low Galactic latitudes are a bit bluer
(which is $J-K_s \sim 0.35$) than that at medium and high Galactic latitudes
(which is between 0.35 and 0.4).  Besides, a middle peak shows up at $J-K_s
\sim 0.55$ at low Galactic latitudes and does not exist in other sky areas.
Several medium and high Galactic latitudes have an extra peak at $J-K_s \sim
0.55$, which is not the same as the middle peak at low Galactic latitudes.

In order to explain the whole sky 2MASS CDs, we create an empirical $M_{K_s}$
against $J-K_s$ HR diagram optimized for the 2MASS data by gathering available
NIR HR diagrams and CMDs \citep{Wainscoat1992, Covey2007,Beletsky2009,
Sarajedini2009, Troisi2010} and incorporate a Milky Way model \citep{Chang2010,
Chang2011}. The HR diagram of the thin disk has a bluer MS-turnoff (i.e.,
younger), and that of the thick disk has a redder MS-turnoff (i.e., older). For
a better explanation of the 2MASS CDs, the result is in favor of a relatively
small scale height thin disk (260 pc, MW-II, compare with 360 pc in MW-I; other
parameters of MW-I and MW-II are given in Table~\ref{model}). We note that the
configuration of MW-II is within the `acceptable' solutions due to the
degeneracy between the structural parameters, which has been pointed out as a
problem for using SC to study the Galactic structure
\citep[e.g.,][]{Chang2011}. Now information from CD helps to rule out the
degeneracy, and we deem that MW-II should be the preferred model.

\begin{figure}
  \epsscale{0.8}
  \plotone{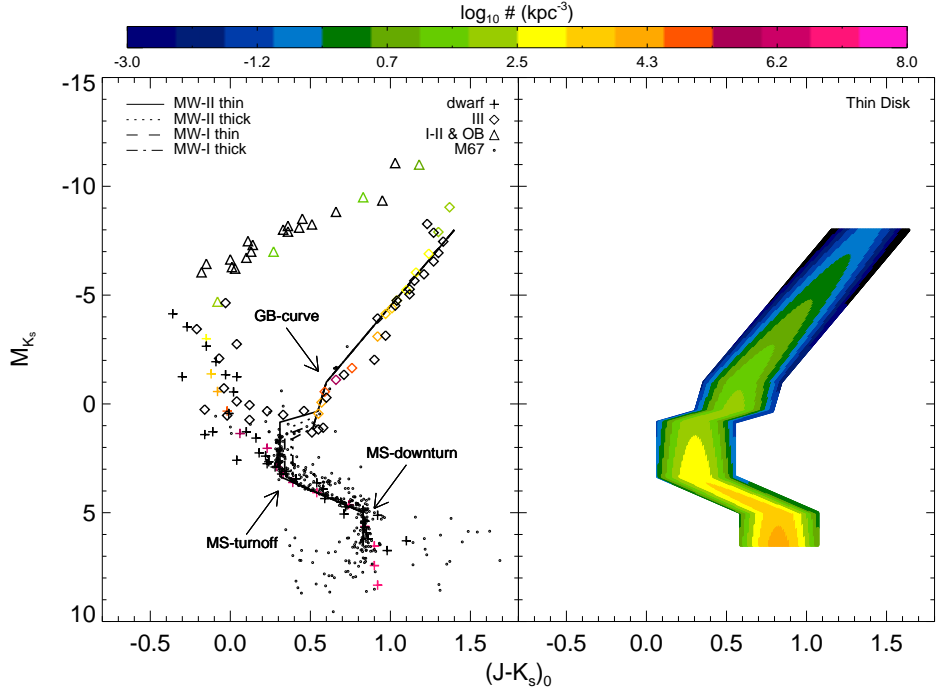}
  \caption{The HR diagram and the Hess diagram. The left panel is the NIR HR diagram. The data points are
  taken from W92 (color symbols), C07 (black symbols) and M67 \citep[small black open circle;][]{Beletsky2009}.
  The upper-right legend indicates luminosity classes and the number density is coded by the color bar.
  C07 does not have number density information, so it is plotted as black.
  Each small black open circle represents one member star of M67. The lines represent different HR diagrams
  used in subsections~\ref{sec2},~\ref{sec3}.
  The upper-left legend gives the corresponding models and components. The right panel is the Hess diagram
  of the thin disk used in MW-II whose number density is color coded.}
  \label{HRD}
  \end{figure}

\begin{figure}
  \epsscale{0.8}
  \plotone{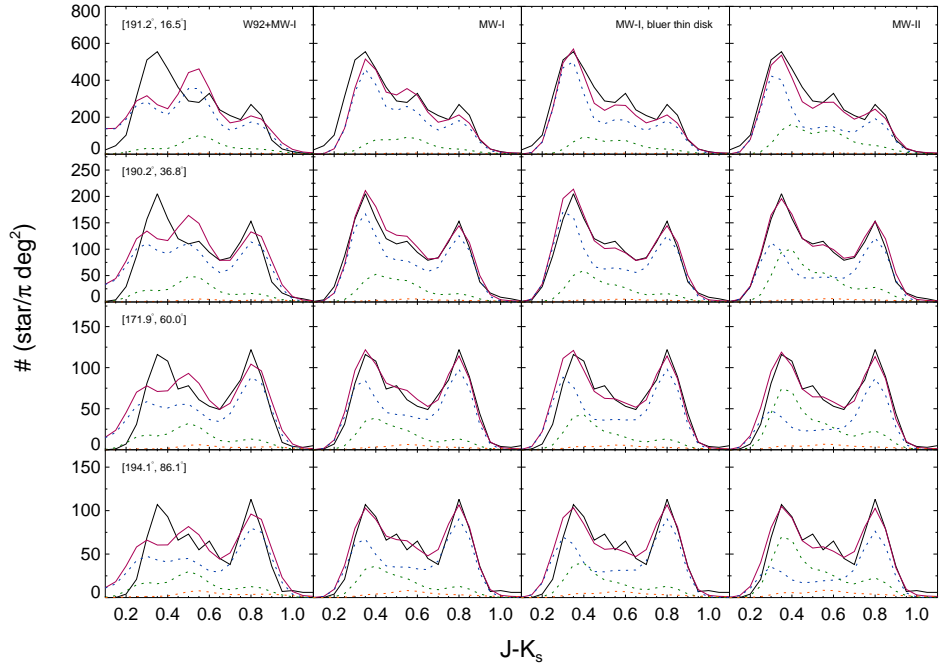}
  \caption{The synthetic CDs of different HR diagrams (see section~\ref{diagram} for details).
  Black line is the 2MASS data, and pink line is the model prediction.
  The blue, green and orange dotted lines represent the thin disk, the thick disk and the halo, respectively.}
  \label{SyntheticCD}
  \end{figure}

We find that the blue peak, the red peak and the middle peak are due to the
MS-turnoff, the MS-downturn and the GB-curve, respectively (see
Fig.~\ref{HRD}). Moreover, the extra peak at some medium or high Galactic
latitudes is due to globular clusters. Our model cannot trace the CDs which
suffer severe extinction and need a more elaborate extinction correction.

When we apply the color-age relation of the MS-turnoff \citep{Sarajedini2009}
to our empirical HR diagram, the thin and thick disks are estimated to be
around 4-5 Gyr and 8-9 Gyr, respectively. The idea is consistent with the Milky
Way's formation theory. We realize that the whole sky 2MASS CDs can be used as
a tool to census the age of stellar populations of the Milky Way in a
statistical manner. To the best of our knowledge, this study is the first
attempt to measure the ages of the thin and thick disks from the whole sky
stellar population.

It is no doubt that the 2MASS PSC provides an unique tool to study the global property
of the Milky Way, not only for its tremendous sky coverage
(which avoids the selection effect of limited sky coverage),
but also for the benefits of using NIR wavelength (which keeps fine angular resolution
and is less affected by interstellar extinction).

\acknowledgements We acknowledge the use of the Two Micron All Sky Survey Point
Source Catalog (2MASS PSC). This work is supported in part by the National
Science Council of Taiwan under the grants NSC-98-2923-M-008-001-MY3 and
NSC-99-2112-M-008-015-MY3.

\begin{deluxetable}{lll}
\tabletypesize{\scriptsize} \tablecaption{The Milky Way model. \label{model}}
\tablewidth{0pt} \tablehead{ \colhead{} & MW-I & MW-II}
\startdata
Density Profile &  &\\
\quad Thin Disk &   &\\
\quad\quad $H_{r1}$ & 3.7 kpc   &  3.7 kpc \\
\quad\quad $H_{z1}$ & 360 pc    &  260 pc   \\
\quad\quad $n_0$    & 0.030 stars/pc$^3$     & 0.039 stars/pc$^3$    \\
\quad\quad $Z_\odot$ & 25 pc    &  \\
\quad\quad $R_\odot$ & 8 kpc \citep{Reid1993}    &  \\
\quad Thick Disk & \\
\quad\quad $H_{r2}$ & 5.0 kpc   & 5.0 kpc \\
\quad\quad $H_{z2}$ & 1020 pc    & 1040 pc \\
\quad\quad $f_2$    & 7\%       & 10 \% \\
\quad Spheroid & \\
\quad\quad $\kappa$ & 0.55   & \\ 
\quad\quad $p$   & 2.6        & \\ 
\quad\quad $f_h$ & 0.20\%     & \\ 
$M_{K_s}$ Luminosity Function & \\
\quad\quad $\gamma$ & 1.85 & \\
\quad\quad $M_b$ & -8 & \\
\quad\quad $M_f$ & 6.5 & \\
Extinction Correction &  & \\
\quad\quad new COBE/IRAS result \citep{Chen1999} & & \\
\quad\quad $E(J-K_s)/E(B-V)$ & 0.53 \citep{Schlegel1998} & \\
\enddata
\end{deluxetable}

\begin{deluxetable}{lll}
\tabletypesize{\scriptsize} \tablecaption{{The Empirical HR Diagram Optimized
for 2MASS.} \label{THRD}} \tablewidth{0pt} \tablehead{ \colhead{Part} &
\colhead{$(J-K_s)_0$} & \colhead{$M_{K_s}$}} \startdata
Common Joints & \\
\quad Faint End of MS & 0.83 & 6.5  \\
\quad MS-Downturn & 0.83  & 5.0 \\
\quad GB-Curve & 0.6 & -1.0  \\
\quad Bright End of Up GB & 1.4 & -8.0  \\
& & \\
Thin Disk & \\
\quad MS-Turnoff & 0.31   & 3.34 \\
\quad Bright End of MS-Turnoff & 0.31   & 0.84 \\
\quad Faint End of Low GB & 0.54 &  0.34   \\
& & \\
Thick Disk & \\
\quad MS-Turnoff  & 0.38   & 3.56\\
\quad Bright End of MS-Turnoff & 0.38   & 1.06 \\
\quad Faint End of Low GB & 0.53 &  0.56   \\
\enddata
\end{deluxetable}

\clearpage

  \begin{figure}
  \epsscale{0.8}
  \plotone{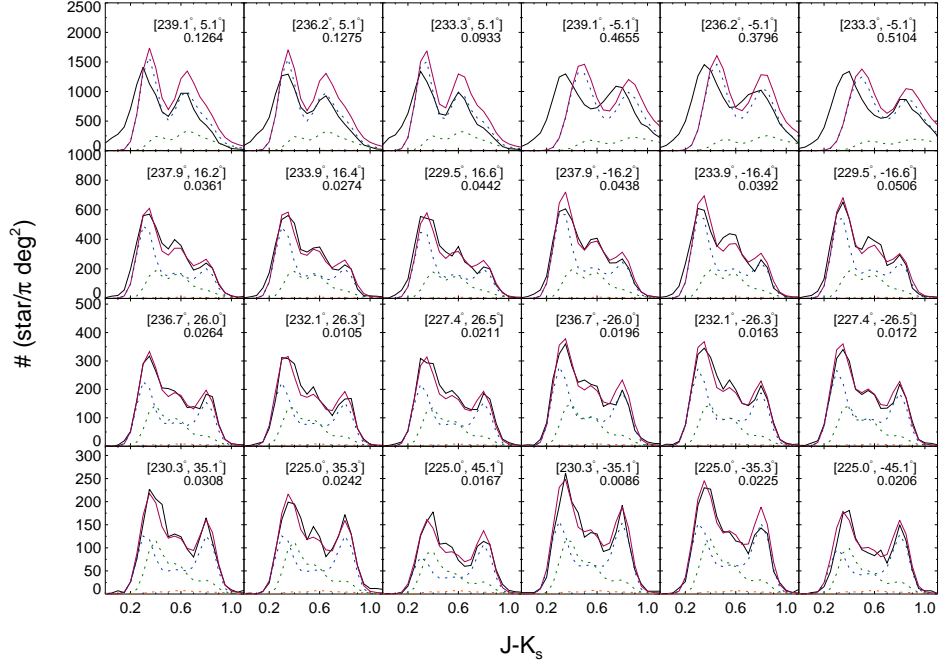}
  \caption{Comparison between 2MASS CDs (black) and model predictions (pink). The coordinates and the
  extinction $E(J-K_s)$ are given at the upper-right corner of each figure.
  The blue, green and orange dotted lines represent the thin disk, the thick disk and the halo, respectively.}
  \label{result1}
  \end{figure}

  \begin{figure}
  \epsscale{0.8}
  \plotone{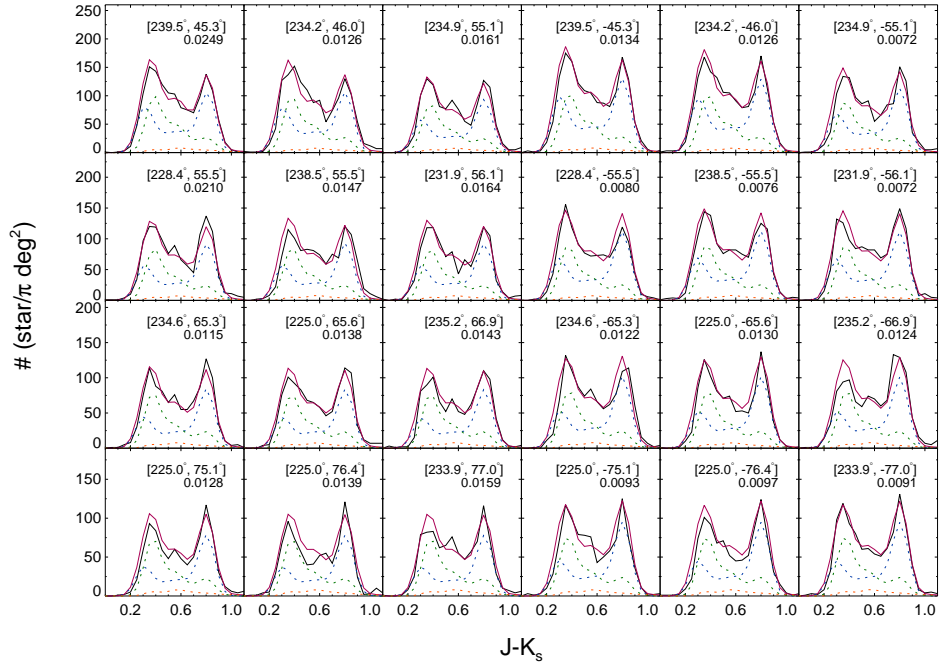}
  \caption{Continuation of Fig.~\ref{result1}.}
  \label{result2}
  \end{figure}

  \begin{figure}
  \epsscale{0.8}
  \plotone{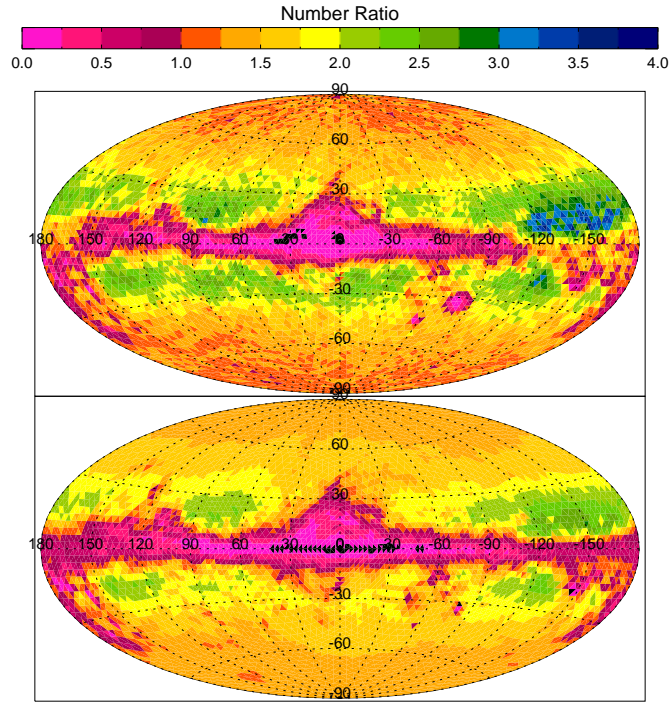}
  \caption{The number ratio of the blue part ($J-K_s < 0.6$) to the red part ($J-K_s > 0.6$).
  Top: 2MASS data. Bottom: model prediction. Note that the very red Galactic disk is the result of
  severe extinction and it does not reflect the intrinsic number ratio in these regions.}
  \label{bridx}
  \end{figure}

  \begin{figure}
  \epsscale{0.8}
  \plotone{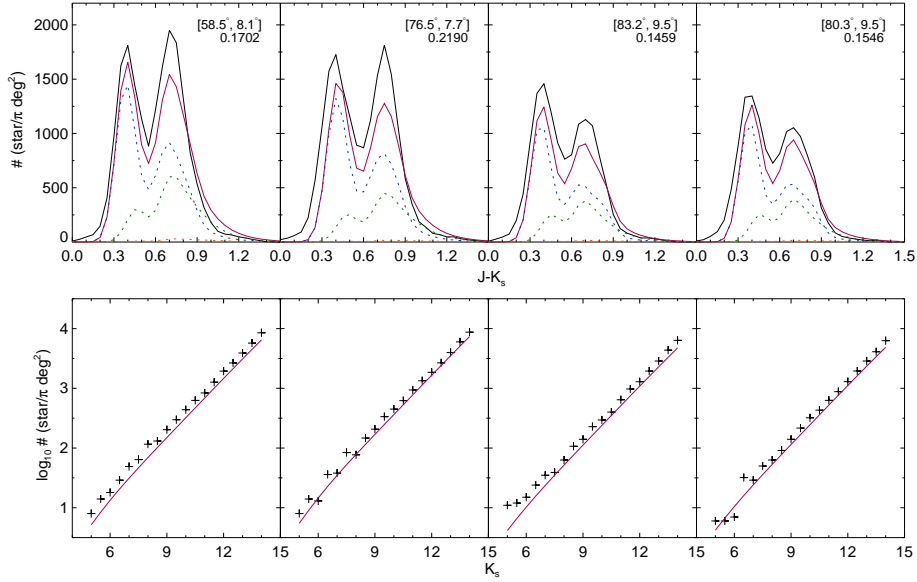}
  \caption{The number inconsistency between the model prediction and the 2MASS CD. The upper panel is the
  comparison of CDs. The upper-right corner shows the coordinates and the extinction $E(J-K_s)$ value.
  The solid-black, solid-pink, dotted-blue, dotted-green and dotted-orange represent
  the 2MASS CD, the model prediction, the thin disk, the thick disk and the halo, respectively.
  The lower panel is the comparison of star counts between 2MASS data (plus)
  and the model prediction (line).}
  \label{inconsist}
  \end{figure}

  \begin{figure}
  \epsscale{0.8}
  \plotone{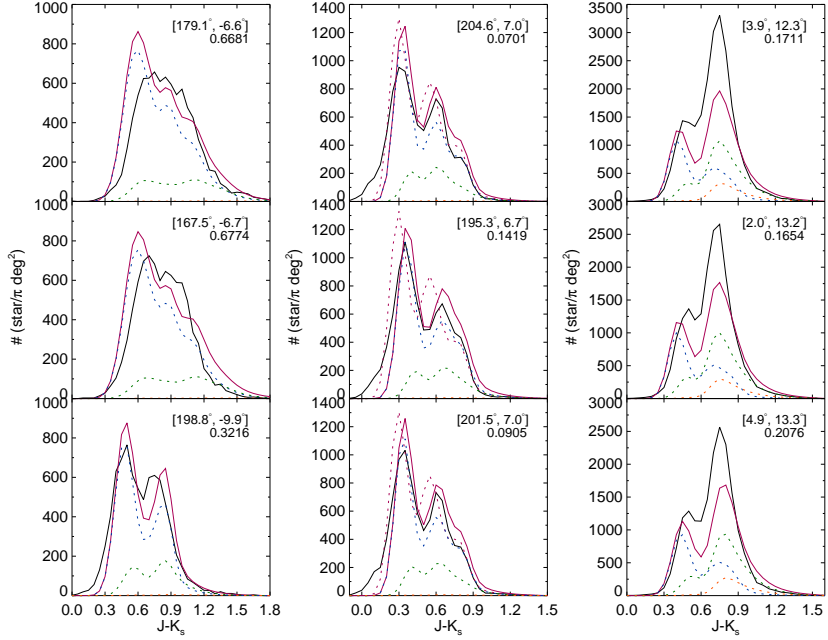}
  \caption{The first column shows the improper extinction correction.
  The second column is the examples of blue population.
  The third column is the examples in Galactic center region.
  The solid-black, solid-pink, dotted-pink, dotted-blue, dotted-green and dotted-orange represent
  the 2MASS CD, the model prediction, the model prediction without extinction correction, the thin disk
  prediction, the thick disk prediction and the halo prediction, respectively.}
  \label{bluep}
  \end{figure}

\end{document}